\documentclass[conference]{IEEEtran}
\IEEEoverridecommandlockouts
\usepackage{amsmath,graphicx}
\usepackage{epstopdf}
\usepackage{setspace}
\usepackage{amsmath}
\usepackage{amssymb}
\usepackage{stfloats}
\usepackage{amsthm}
\usepackage{subcaption}
\usepackage{amsfonts} 
\usepackage{color}
\usepackage{xcolor}

\usepackage{graphicx}

\usepackage[flushleft]{threeparttable} 

\usepackage{upgreek}
\usepackage{bm}
\usepackage{nicefrac}
\usepackage{subcaption}

\usepackage{multirow}
\usepackage{cite}
\usepackage{mathtools}
\usepackage{stmaryrd}
\usepackage[mathscr]{euscript}
\usepackage{array}
\usepackage{mathrsfs}

\usepackage{soul}
\newtheorem{remark}{Remark}

\newtheorem{theorem}{Theorem}

\begin{document}

 \title{A Non-Uniform Quantization Framework
for Time-Encoding Machines 
\thanks{$^*$Equal contribution.
}}%

\author{\IEEEauthorblockN{Kaluguri Yashaswini*, Anshu Arora* and Satish~Mulleti}
\IEEEauthorblockA{\textit{Department of Electrical Engineering,  Indian Institute of Technology Bombay},
Mumbai, India\\
Emails: yashkaluguri2005@gmail.com, anshuarora2604@gmail.com, mulleti.satish@gmail.com}}

%
%
          

%
\maketitle

\noindent \begin{abstract}
Time encoding machines (TEMs) provide an event-driven alternative to classical uniform sampling, enabling power-efficient representations without a global clock. 
While prior work analyzed uniform quantization (UQ) of firing intervals, we show that these intervals are inherently non-uniformly distributed, motivating the use of non-uniform quantization (NUQ). 
We derive the probability distribution of firing intervals for a class of bandlimited signals and design a power-law-based NUQ scheme tailored to this distribution. 
Simulations demonstrate that NUQ significantly outperforms UQ under the same bit budget. 
We also compare TEMs with non-uniform sampling (NUS), where both amplitudes and timings require quantization, and show that TEM--NUQ achieves lower error at half the transmission cost. 
These results highlight the advantages of distribution-aware quantization and establish TEM--NUQ as an efficient alternative to conventional UQ and NUS schemes.
\end{abstract}


\section{INTRODUCTION}
\noindent Sampling and quantization are fundamental to digital signal processing, enabling continuous-time signals to be represented, stored, and transmitted efficiently. 
Quantization is typically uniform (UQ) or non-uniform (NUQ). 
While UQ is simple, it is suboptimal when the signal amplitudes are not uniformly distributed, whereas NUQ adapts resolution to the statistics of the data and yields improved rate–distortion performance \cite{gersho2012vector, companding, max_lloyd, hila2022time, escort1}.  

As in the quantization, unlike uniform sampling \cite{jerri1977shannon}, event-driven strategies such as non-uniform sampling (NUS) \cite{marvastinonuniform, irregsampling, aldroubi2001nonuniform} and time encoding machines (TEMs) \cite{lazar2005time, gontier2014sampling, VBIFTEM, omar2024adaptive, hilaTEM} offer more efficient representations by adapting sampling locations to the signal variations. 
In particular, the integrate-and-fire TEM (IF--TEM) stands out for its simplicity and low power consumption \cite{PerfectRecovery, compressediftem, lowpower, mulleti2023powerefficientsampling, hilton2021guaranteed}. 
By integrating the input and recording firing times whenever a threshold is reached, the IF--TEM produces time encodings whose inter-event intervals adapt naturally to the input dynamics.  

Prior work \cite{PerfectRecovery} analyzed UQ of IF--TEM intervals and showed that the induced quantization error has the same upper bound as quantizing amplitudes in NUS. 
However, since the firing intervals are rarely uniformly distributed, UQ is inefficient for IF--TEM outputs. 
This motivates the design of NUQ schemes tailored to the statistics of time encodings.

In this paper, we develop a statistical framework and quantization strategies for IF--TEM encodings. 
Our main contributions are:
\begin{itemize}
    \item We derive the probability distribution of firing intervals for a class of stationary, ergodic, bandlimited processes, relating it explicitly to the amplitude distribution. 
This analysis highlights that the induced distributions are far from uniform, thereby demonstrating the necessity of NUQ for time encodings.
    \item Using this characterization, we design a power-law-based non-uniform quantizer that adapts to the distribution of firing intervals and achieves lower reconstruction error than UQ for the same bit budget.  
    \item We compare quantization in IF--TEM with that in NUS, where both times and amplitudes must be quantized, and demonstrate that TEM--NUQ achieves superior accuracy at a lower communication cost.  
\end{itemize}
Together, these contributions establish that exploiting the statistical structure of firing intervals enables more efficient quantization and positions TEMs as a practical alternative to conventional 
uniform sampling and quantization schemes.

In the following Section~\ref{sec:ProblemFormulation}, we formulate the problem. In Section~\ref{sec:NUQdesign}, we derive the distribution of time encodings, as well as the design of a nonuniform quantizer from a theoretical and practical standpoint. We elaborate on our results in Section~\ref{sec:simresult}, and conclude in Section~\ref{sec:conclusion}.

\section{PROBLEM FORMULATION}
\label{sec:ProblemFormulation}
\noindent We consider the input signal as a real-valued, stationary, ergodic, and bandlimited stochastic process 
$\{f(t)\}_{t \in \mathbb{R}} \in \mathcal{B}_{\Omega_0,c}$, 
with spectral support in $[-\Omega_0,\Omega_0]$ and bounded amplitude, that is, 
$|f(t)| \leq c$. For any $f(t) \in \mathcal{B}_{\Omega_0, c}$, its Fourier 
transform is defined as $F(\Omega) = \int_{-\infty}^{\infty} f(t) e^{\mathrm{j}\Omega t} \, \mathrm{d}t,$
and satisfies $F(\Omega) = 0$ for all $|\Omega| > \Omega_0$. In addition, the 
amplitude of these signals is bounded such that $|f(t)| \leq c$. It is 
well-known by the Shannon-Nyquist sampling theory \cite{nyquist1928certain},\cite{jerri1977shannon} that any $f(t) \in \mathcal{B}_{\Omega_0, c}$ can be exactly reconstructed from its uniform samples $\{f(nT_s)\}$, provided that the sampling period obeys $T_s < T_{Nyq} = \frac{\pi}{\Omega_0}.$
The discrete representation of the samples $f(nT_s)$ is then quantized for further processing. Alternatively, the signal can also be reconstructed from the NUS set $\{f(t_n), t_n\}_{n \in \mathbb{Z}}$ provided that $t_{n+1} - t_n < T_{Nyq}$ \cite{feichtinger1994theory}. However, unlike the uniform sampling, here, both amplitudes and sampling locations have to be quantized.  

\begin{figure}[!t]
        \centering
        \includegraphics[width= 2.5 in]{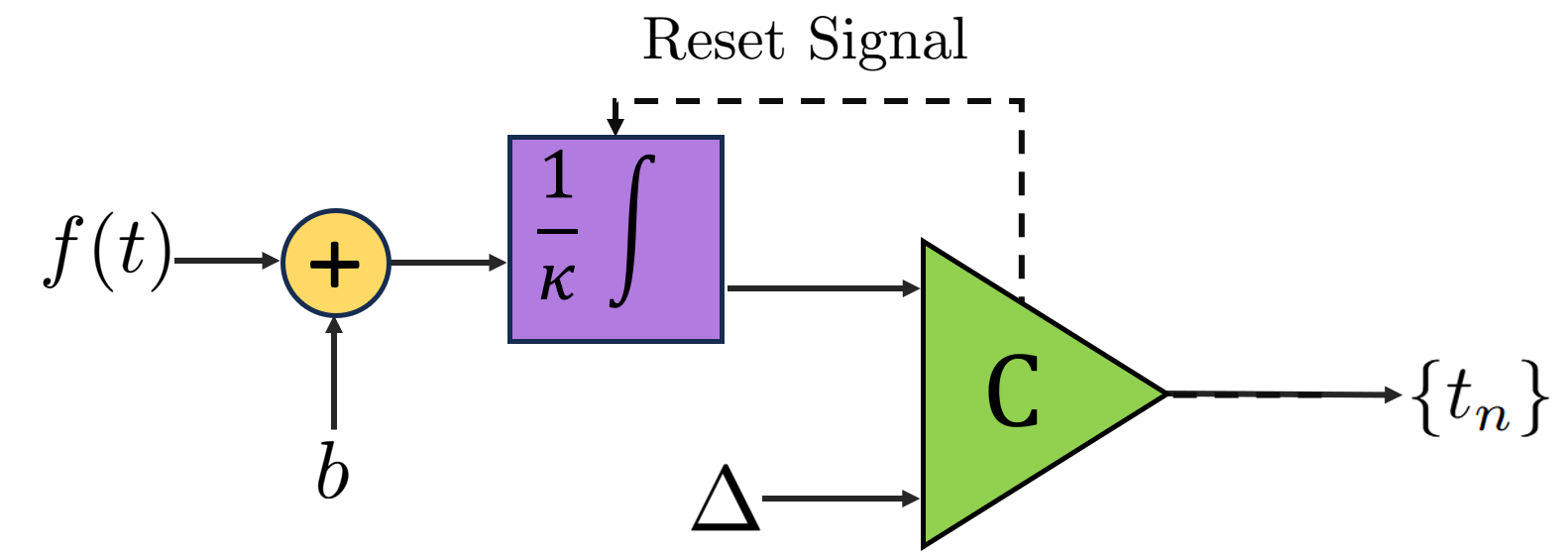}
        \caption{Schematic of the IF--TEM}
        \label{fig:iftem_block}
    \end{figure}

In this work, we consider IF--TEM for the discrete representation of signals in $\mathcal{B}_{\Omega_0, c}$. In a conventional IF--TEM as illustrated in Fig.~\ref{fig:iftem_block}, the input signal $f(t) \in \mathcal{B}_{\Omega_0, c}$ is first biased to obtain a strictly non-negative signal 
$f(t)+ b$, where $b > c$. This biased signal is then integrated 
with a scaling factor of $\tfrac{1}{\kappa}$. 
A firing event occurs when the output of the integrator reaches a fixed threshold 
$\Delta$, after which the integrator is reset. This process generates a strictly 
increasing sequence of firing instants 
\[
\cdots < t_{n-1} < t_n < t_{n+1} < \cdots,
\]
which are related to the signal through the condition
\begin{align}
    \frac{1}{\kappa} \int_{t_{n-1}}^{t_n} \big(f(t)+b\big) \, \mathrm{d}t = \Delta.
    \label{eq:iftem1}
\end{align}
In a nutshell, the time encodings $\{t_n\}$ are the discrete representations of the signal $f(t)$, unlike the uniform samples $\{f(nT_s)\}$ or the NUS set $\{t_n, f(t_n)\}$. The time differences, 
\begin{align}
    T_n = t_{n+1} - t_n,
\end{align}
are bounded as
\begin{align}
    \frac{\kappa \Delta}{b + c} \leq T_n \leq \frac{\kappa \Delta}{b - c}. \label{eq:tn_bound}
\end{align}
The bounds are derived by using  \eqref{eq:iftem1}, and noting that the dynamic range of the signal is $[-c, c]$.

Once the intervals $\{T_n\}$ are given, the signal $f(t)$ can 
be perfectly reconstructed by applying an iterative 
algorithm, provided that $T_n < T_{\text{Nyq}}, \ \forall n$ \cite{PerfectRecovery}. We must select $\kappa$ and $\Delta$ according to \eqref{eq:tn_bound} such that this condition is satisfied. Without loss of generality, we set $\kappa = 1$ and keep $\Delta$ as a free variable.

In practice, the quantities 
$T_n$s necessitate quantization for subsequent processing and storage. Quantization leads to an error in reconstruction, which can be reduced by either increasing the number of bits or by designing a NUQ that matches the distribution of the time encodings. This, in turn, depends on the distribution of the signal's amplitude and the IF--TEM parameters. This necessitates establishing a relation between the two distributions and then designing a suitable NUQ for $T_n$s. We consider this problem and present our results in the following section.

 \begin{remark}[Deterministic vs. stochastic modeling]
The IF--TEM principle described above applies to any bounded, bandlimited signal, 
and the reconstruction guarantees hold in a purely deterministic setting. 
However, when analyzing the effect of quantization, it becomes essential to 
characterize the \emph{statistics} of the quantities being quantized. 
For this reason, in the following section, we adopt a stochastic model, 
treating $\{f(t)\}_{t \in \mathbb{R}}$ as a stationary, ergodic, bandlimited random process 
with an amplitude distribution. 
This allows us to rigorously define the induced distribution of the firing intervals $\{T_n\}$ 
and to design a non-uniform quantizer tailored to these statistics. 
Through ergodicity, these ensemble distributions can be accurately approximated 
from single realizations, thereby bridging the deterministic and stochastic viewpoints.
\end{remark}

\section{DISTRIBUTION OF TIME ENCODINGS \& NUQ DESIGN}
\label{sec:NUQdesign}
\noindent In this section, we first derive the distribution of the time encodings and then discuss the NUQ design aspects.

\subsection{\textbf{Theoretical Results}}
\noindent Efficient non-uniform quantization requires knowledge of the underlying statistics of the data being quantized. 
In our setting, these statistics are determined by the amplitude law of the input process, 
which we denote by the \emph{ensemble distribution} to distinguish it from empirical histograms obtained from realizations.  

For any fixed $t \in \mathbb{R}$, let $\mathbf{F}$ be the random variable corresponding to the amplitude $f(t)$, 
with ensemble law $P_{\mathbf{F}}(f)$.  
Equivalently, at any time instant $t$, the random variable $f(t)$ is distributed according to $P_{\mathbf{F}}$. 

The IF--TEM produces a sequence of time encodings $\{T_n\}$ according to \eqref{eq:iftem1}. 
We denote by $\mathbf{T}$ the random variable governed by the distribution $P_{\mathbf{T}}(T)$, 
which is induced from $P_{\mathbf{F}}(f)$ through the IF--TEM mapping. 
This relationship is formalized below.

\begin{theorem}[Ensemble distribution of the time encodings]
Let $\{f(t)\}_{t \in \mathbb{R}}$ be a stationary, ergodic, bandlimited stochastic process with bounded amplitude 
$|f(t)| \le c$ almost surely, and with amplitude distribution $P_\mathbf{F}(f)$ supported on $[-c,c]$. 
Then the ensemble distribution of the time encodings $\{T_n\}$ generated by the IF--TEM is
\begin{align}
    P_{\mathbf{T}}(T) =
    \begin{cases}
        P_{\mathbf{F}}\!\left(\dfrac{\Delta}{T} - b\right)\,\dfrac{\Delta}{T^{2}}, 
        & \dfrac{\Delta}{b+c} \leq T \leq \dfrac{\Delta}{b-c}, \\[1ex]
        0, & \text{otherwise},
    \end{cases}
    \label{eq:transformation}
\end{align}
where $c$ is the maximum amplitude of the process.
\label{th:distribution}
\end{theorem}

\begin{proof}
The IF--TEM encoding law is
\begin{align}
    \int_{t_n}^{t_{n+1}} \big(f(t) + b \big)\, dt = \Delta,
    \label{eq:iftem_integral_law}
\end{align}
which can be equivalently written as
\begin{align}
  T_n = \frac{\Delta}{\left( \frac{1}{t_{n+1}-t_{n}} \int_{t_n}^{t_{n+1}} f(t)\, dt \right) + b}.
  \label{eq:Tn_expression}
\end{align}
Since $f(t)$ is bandlimited, it is continuous, and by the Mean Value Theorem for integrals~\cite[Chapter~6]{rudin1976principles}, 
there exists a point $\tau_n \in [t_n, t_{n+1}]$ such that
\begin{align}
    \frac{1}{t_{n+1}-t_n}\int_{t_n}^{t_{n+1}} f(t)\, dt = f(\tau_n).
    \label{eq:mvt}
\end{align}
Substituting into \eqref{eq:Tn_expression} gives
\begin{align}
    T_n = \frac{\Delta}{f(\tau_n) + b}.
    \label{eq:Tn_mvt}
\end{align}
By stationarity, the distribution of $f(\tau_n)$ coincides with the ensemble distribution of $f(t)$ at any fixed time~\cite[Chapter~9]{papoulis2002probability}. Thus,
\begin{align}
    f(\tau_n) \sim P_{\mathbf{F}}, \qquad T_n \sim P_{\mathbf{T}}.
\end{align}
The probability density function (PDF) of $\mathbf{T}$ induced by the mapping $g(f)=\Delta/(f+b)$ is obtained via the change-of-variables theorem~\cite[Chapter~5]{papoulis2002probability}. 
Since $g$ is monotone on $[-c,c]$, we have
\begin{align}
    P_{\mathbf{T}}(T) 
    &= P_{\mathbf{F}}\!\big(g^{-1}(T)\big) \, \bigg|\frac{d}{dT} g^{-1}(T)\bigg|,
    \label{eq:change_of_vars}
\end{align}
with
\begin{align}
    g^{-1}(T) = \frac{\Delta}{T} - b, 
    \qquad
    \frac{d}{dT} g^{-1}(T) = -\frac{\Delta}{T^2}.
\end{align}
A substitution gives
\begin{align}
    P_{\mathbf{T}}(T) = P_{\mathbf{F}}\!\left(\frac{\Delta}{T} - b\right)\,\frac{\Delta}{T^2}.
    \label{eq:transform_pdf_final}
\end{align}
Finally, since ${\mathbf{F}} \in [-c, c]$, the support of ${T}$ is
\begin{align}
    \frac{\Delta}{b+c} \le T \le \frac{\Delta}{b-c}.
    \label{eq:support_T}
\end{align}
Equations \eqref{eq:transform_pdf_final} and \eqref{eq:support_T} together establish the claimed distribution.
\end{proof}

\begin{remark}[Empirical vs. Ensemble Distribution]
The functional form of $P_{\mathbf{T}}(T)$ provides the ensemble distribution of firing intervals. 
In practice, however, only a single realization of $\{T_n\}$ is observed.  

Since $\{f(\tau_n)\}_{n \in \mathbb{N}}$ is stationary and ergodic, and the mapping 
$g(x) = \tfrac{\Delta}{b+x}$ is measurable and bounded (as ensured by~\eqref{eq:support_T}), 
it follows that the sequence $\{T_n\}$ is also stationary and ergodic \cite[Chapter~6]{walters1982introduction}.  

Thus, by Birkhoff’s Ergodic Theorem \cite[Chapter~1]{walters1982introduction}, for any integrable function $\varphi$,
\begin{align}
    \frac{1}{N}\sum_{n=1}^N \varphi(T_n) 
    \xrightarrow[N \to \infty]{\text{a.s.}} 
    \mathbb{E}[\varphi(\mathbf{T})].
\end{align}
In particular, for $\varphi =\mathbf{1}_A$ and the Borel set $A \subseteq [\Delta/(b+c), \, \Delta/(b-c)]$,
\begin{align}
    \frac{1}{N}\sum_{n=1}^N \mathbf{1}_{\{T_n \in A\}}
    \xrightarrow[N \to \infty]{\text{a.s.}} 
    P_{\mathbf{T}}(A).
\end{align}
Therefore, the empirical histogram of $\{T_n\}$ converges almost surely to the theoretical law $P_{\mathbf{T}}$.  
This ensures that a non-uniform quantizer designed from the ensemble law is valid in practice based on long single realizations.  
\end{remark}

\begin{figure}[!t]
    \centering
    \begin{subfigure}[b]{0.48\linewidth}
        \centering
        \includegraphics[width=\linewidth]{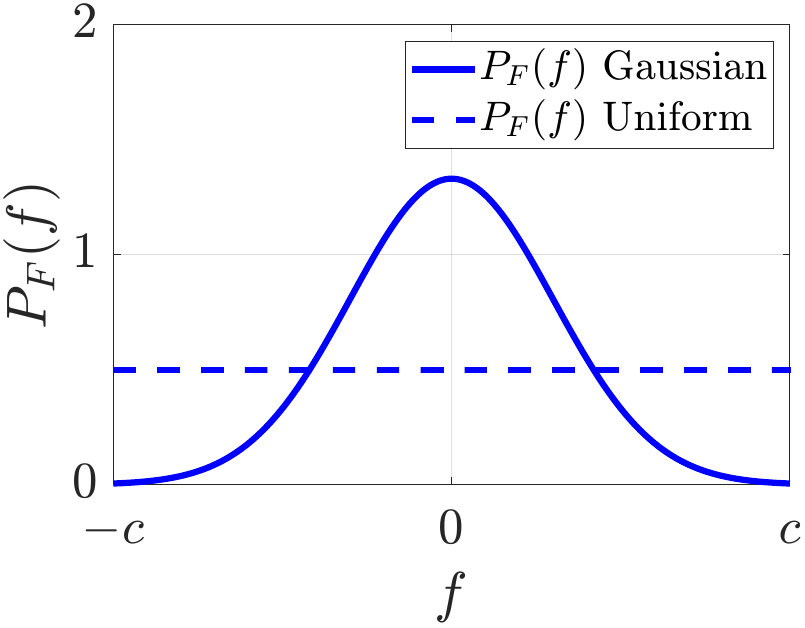}
        \caption{Truncated Gaussian signal distribution, along with uniform amplitude distribution.}
        \label{fig:signal_gaussian}
    \end{subfigure}
    \hfill
    \begin{subfigure}[b]{0.48\linewidth}
        \centering
        \includegraphics[width=\linewidth]{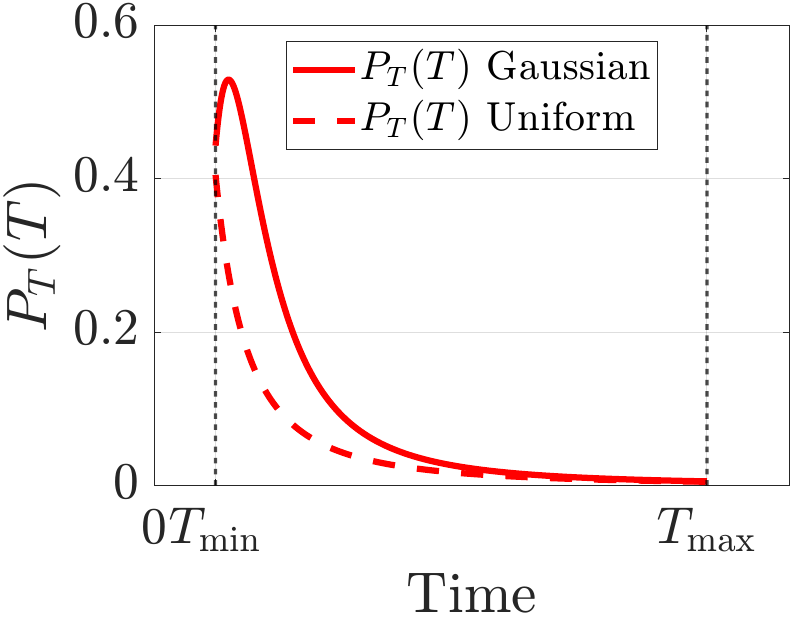}
        
        \caption{For corresponding, Gaussian and uniform inter-event distributions, time PDFs $P_{\mathbf{T}}(T)$.}
        \label{fig:time_gaussian}
    \end{subfigure}
  \caption{Relationship between the signal distribution and the induced firing time distribution in the IF--TEM framework.}
    \label{fig:gaussian_distributions}
\end{figure}

In Fig.~\ref{fig:gaussian_distributions}, we compare the distributions of signal amplitudes and the corresponding firing intervals obtained via \eqref{eq:transformation} for Gaussian and uniform amplitude laws. 
A key observation is that the nonlinear mapping of amplitudes to firing intervals inherently distorts the distribution: even when the amplitudes are uniformly distributed, the resulting time encodings are far from uniform. 
For Gaussian amplitudes, the effect is more pronounced, with firing intervals clustering in specific regions rather than spreading evenly. 
These concentration patterns make uniform quantization fundamentally inefficient, as it fails to match the true statistical structure of the data. 
This motivates the use of non-uniform quantization, which allocates finer resolution where encodings are more likely to occur, thereby reducing distortion and improving reconstruction performance.

\subsection{\textbf{Practical NUQ Design for IF--TEM}}
\begin{figure}[!t]
    \begin{subfigure}[b]{0.48\linewidth}
        \centering
        \includegraphics[width=\linewidth]{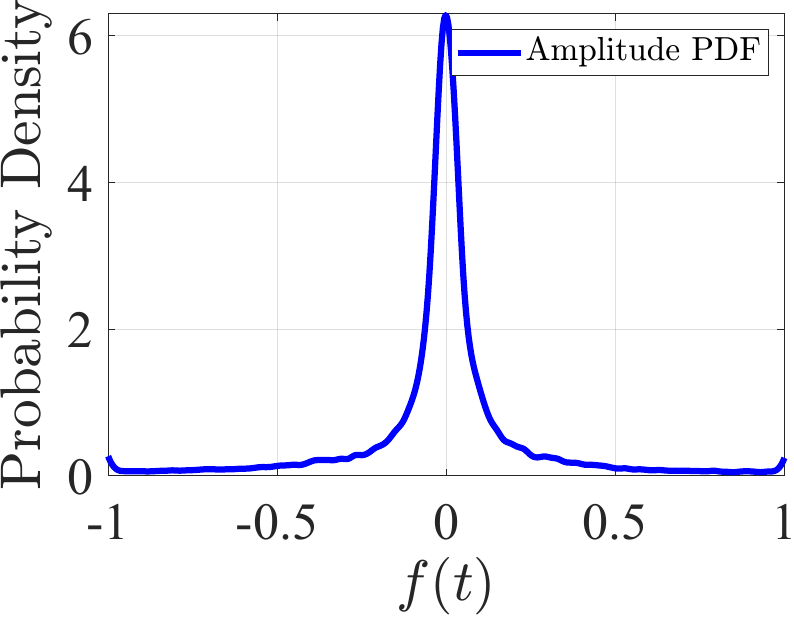}
        \caption{Amplitude distribution from simulated signals.}
        \label{fig:amplitude_results}
    \end{subfigure}
    \hfill
    \begin{subfigure}[b]{0.48\linewidth}
        \centering
        \includegraphics[width=\linewidth]{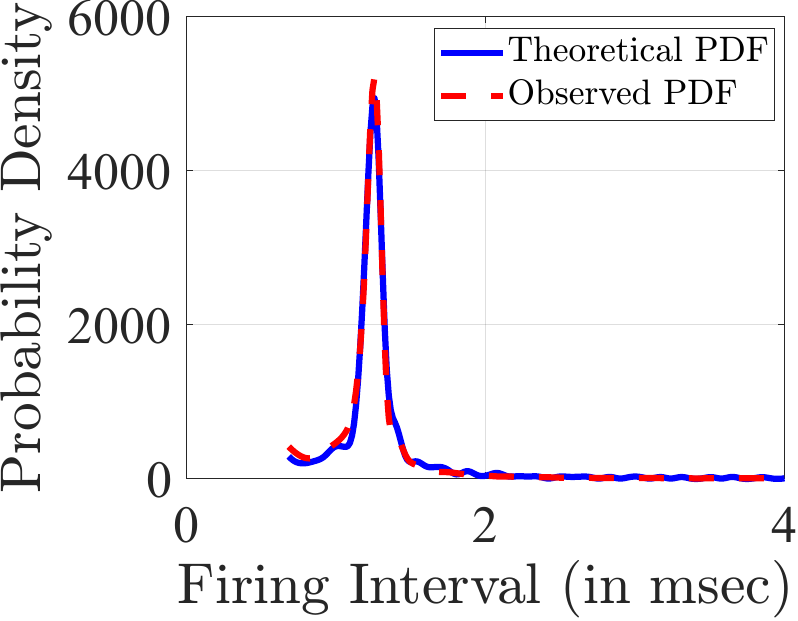}
        \caption{Observed and theoretical distributions of firing times.}
        \label{fig:histogram_counts}
    \end{subfigure}
\caption{A comparison of the empirically observed distribution and the theoretical one. 
\eqref{eq:transformation} to the PDF in (a). } 
\label{Fig:expected_vs_actual}
    \end{figure}

\noindent To design a NUQ, we need exact knowledge of the distributions of amplitudes. In practice, only the empirical distributions can be determined. Before discussing NUQ choices, we first assess the closeness of the empirical distributions to the true ones.

To this end, in Fig.~\ref{Fig:expected_vs_actual}, we illustrate the distributional amplitude and time encodings of the IF--TEM outputs. 
To generate these results, $100$ independent realizations of a bandlimited stochastic process were synthesized as sums of sinc functions with spectral support up to $50\,\text{Hz}$. 
Each signal was sampled at a resolution of $1\,\mu\text{s}$ to obtain high-fidelity approximations of the underlying distributions, 
where the continuous-time, bandlimited nature of the process ensures accurate representation of the theoretical models.  

Fig.~\ref{Fig:expected_vs_actual}(a) shows the averaged histogram of the signal amplitudes computed over the $100$ independent realizations(signals). 
This averaging procedure provides an empirical estimate of the ensemble law $P_{\mathbf{F}}(f)$~\cite[chapter 7]{papoulis2002probability} of the bandlimited stochastic process.  Fig.~\ref{Fig:expected_vs_actual}(b) presents an overlay of the theoretically derived PDF of the interspike intervals $P_{\mathbf{T}}(T)$, obtained from \eqref{eq:transformation}, 
and the observed distribution of $\{T_n\}$ estimated from the simulated signals as for the amplitudes. 
The close agreement between the theoretical and simulated distributions validates the analytical derivation of the interspike interval ensemble distribution and confirms the accuracy of the IF--TEM simulations.

Motivated by this characterization, we now design non-uniform quantizers for the firing intervals. 
In principle, NUQ can be implemented directly using optimization-based methods such as the Lloyd–Max algorithm, 
or indirectly by applying a suitable transformation followed by uniform quantization. 
The latter approach is particularly well-suited for the present application, 
since power-law based companding schemes are inherently more robust to distribution mismatches and outliers than direct Lloyd–Max designs. 
Accordingly, we employ a companding-based strategy similar to~\cite{escort1, escort2}, in which the distribution is first nonlinearly transformed as  
\begin{align}
    P_\mathbf{T}(T) 
    &\;\mapsto\; 
    \frac{P_\mathbf{T}(T)^{\gamma}}
         {\int P_\mathbf{T}(u)^{\gamma}\,du},
    \quad 0 < \gamma < 1.
\end{align}

With $\gamma \in (0,1)$, this transformation flattens concentrated regions of the distribution and expands underrepresented regions, 
thereby facilitating more balanced quantization. 
Uniform quantization is then performed in the transformed domain, and the outputs are mapped back to yield the quantized sequence $\{\hat{T}_n\}$. 
These quantized encodings are subsequently used in the reconstruction stage to estimate $f(t)$.

\begin{figure}[!t]
    \centering
    \includegraphics[width=0.8\linewidth]{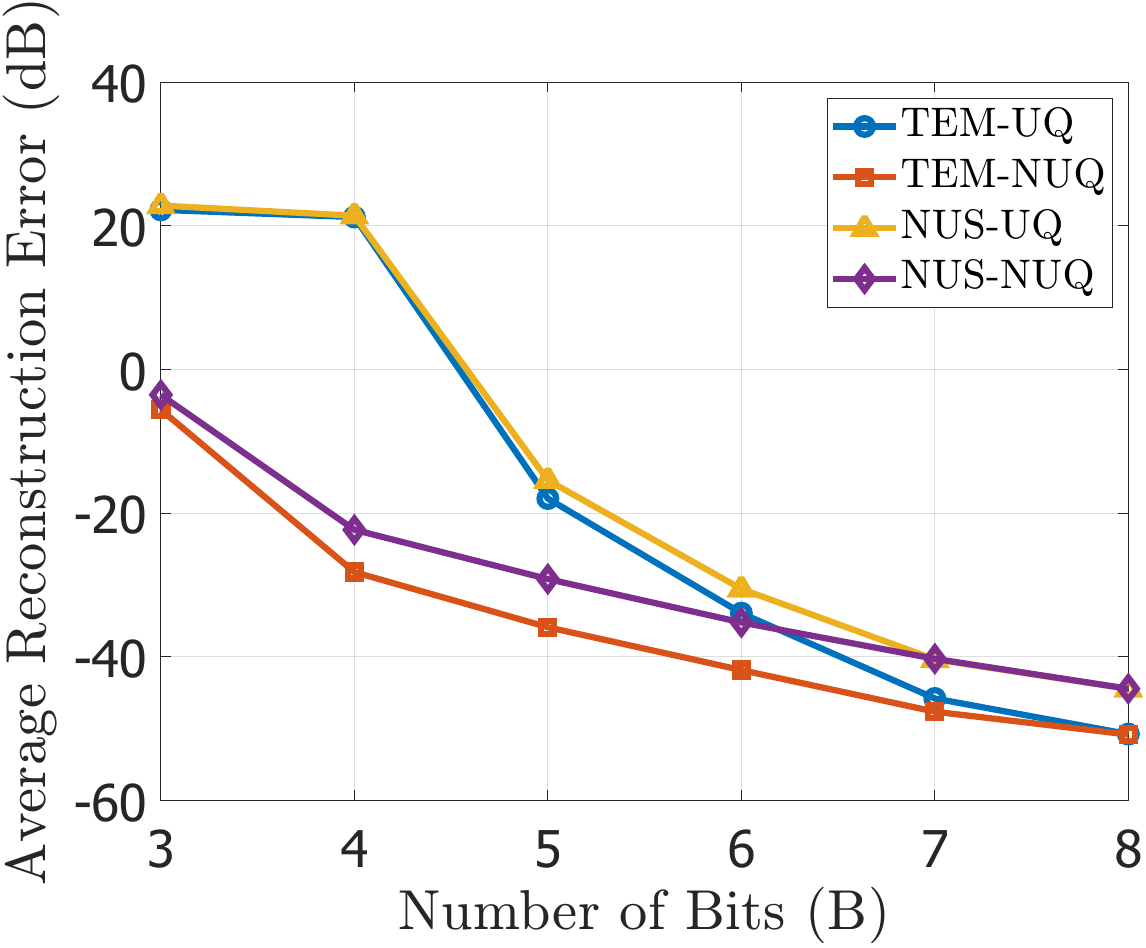}
    \caption{NMSE comparison of TEM and NUS under uniform and non-uniform quantization.}
    \label{fig:main_figure}
\end{figure}

 \begin{figure}[!t]
    \centering
    \begin{subfigure}{0.48\linewidth}
        \centering
        \includegraphics[width=\linewidth]{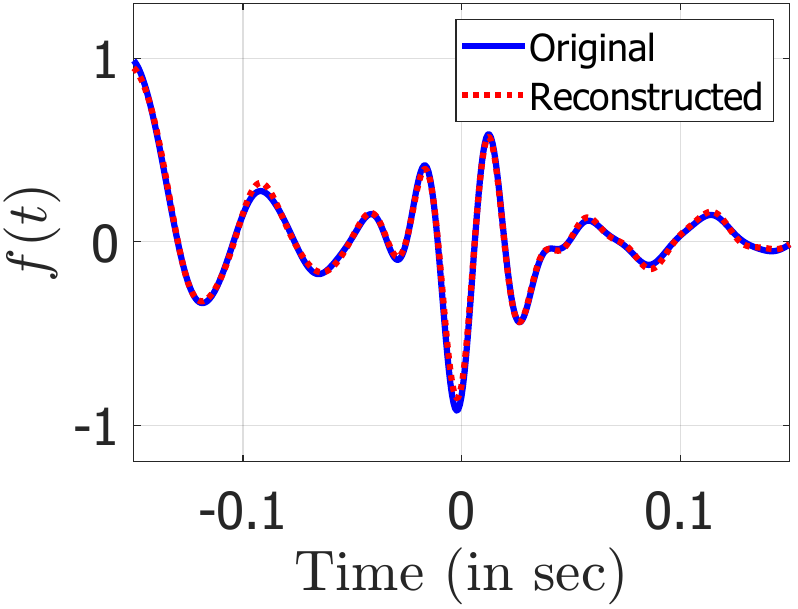}
        \caption{Non-Uniform quantization in NUS with 4 bits ($\text{NMSE} = -23.95 \,\mathrm{dB}$).}
        \label{fig:TEM_error_3}
    \end{subfigure}
    \hfill
    \begin{subfigure}{0.48\linewidth}
        \centering
        \includegraphics[width=\linewidth]{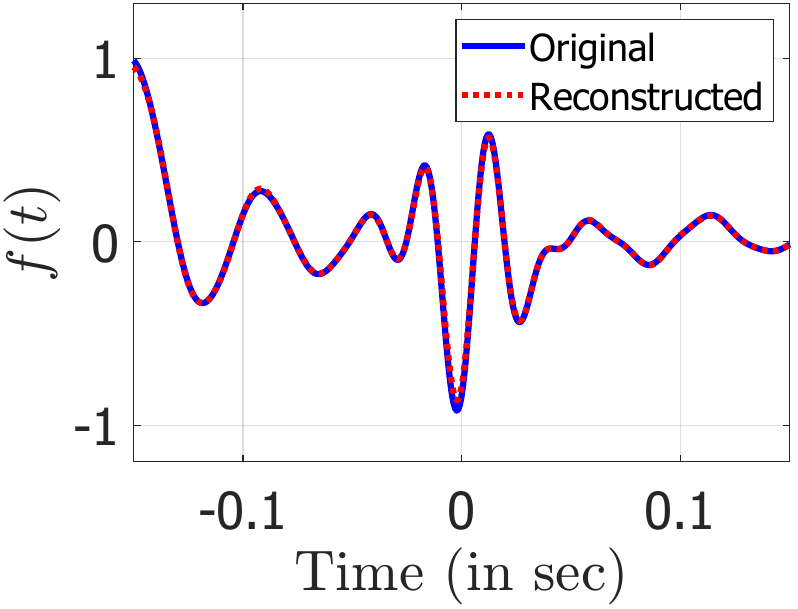}
        \caption{Non-Uniform  quantization in TEM with 4 bits ($\text{NMSE} = -29.8 \,\mathrm{dB}$).}
        \label{fig:TEM_error_4}
    \end{subfigure}
    \caption{Reconstruction error comparison for IF--TEM and NUS under non-uniform quantization of time encodings.}
    \label{fig:TEM_error_time_UQ}
\end{figure}

\section{SIMULATION RESULTS}
\label{sec:simresult}
\noindent For simulations, we generated signals as sums of sinc functions with spectral support up to $50$~Hz and amplitude range $[-1,1]$, restricted to the interval $[-0.45 s,0.45s]$.  
For the TEM, we set the bias $b=1.2$ and threshold $\Delta=0.0015$ such that the perfect reconstruction guarantees are preserved.

For comparison, we also implemented a non-uniform sampling (NUS) scheme, where samples $\{f(t_n),t_n\}$ were taken at the IF--TEM firing instants, ensuring adaptivity similar to TEM. 
In both IF--TEM and NUS, the time encodings were quantized using a power-law-based compander. Across all experiments, we found that setting $\gamma = 0.1$ yielded the lowest error. In NUS, the amplitudes $f(t_n)$ were quantized using an NUQ designed via the Lloyd--Max algorithm, which consistently yields the best results in this setting.

Fig.~\ref{fig:main_figure} presents the primary comparison. 
For NUS, the notation $B$ bits indicates that $B$ bits are used for time encodings and an additional $B$ for amplitudes, giving a total of $2B$ bits---twice the communication cost of TEM. 
The results show that NUQ consistently outperforms UQ in both IF--TEM and NUS. 
Moreover, TEM--NUQ achieves lower error than NUS--NUQ, despite using half the number of bits. 
Thus, at higher bit rates, TEM--NUQ simultaneously improves efficiency and reconstruction accuracy.

Finally, Fig.~\ref{fig:TEM_error_time_UQ} visualizes reconstruction quality. 
With $-30$~dB NMSE as the threshold for acceptable reconstruction, this level of accuracy is achieved with 4 bits in TEM--NUQ, 6 bits in TEM--UQ, 10 bits in NUS--NUQ, and 12 bits in NUS--UQ. 
This highlights the superior rate--distortion tradeoff offered by the proposed TEM--NUQ framework.

\section{CONCLUSION}
\label{sec:conclusion}
\noindent We presented a framework for non-uniform quantization of IF--TEM encodings by exploiting the statistical relation between signal amplitudes and firing intervals. 
Our analysis established the induced distribution of the intervals and motivated a power-law-based NUQ design, which consistently outperforms uniform quantization. 
Comparisons with NUS further showed that TEM--NUQ achieves lower reconstruction error at half the bit rate, demonstrating both accuracy and efficiency. 
Overall, these results highlight the benefits of incorporating statistical knowledge into quantizer design and position TEMs as a compelling alternative to conventional sampling and quantization schemes.

\bibliographystyle{IEEEtran}
\bibliography{ref_icassp}

\section{APPENDIX}
\subsection{Comparison to Uniform Quantization}
\begin{figure}[!t]
    \centering
    \includegraphics[width=0.8\linewidth]{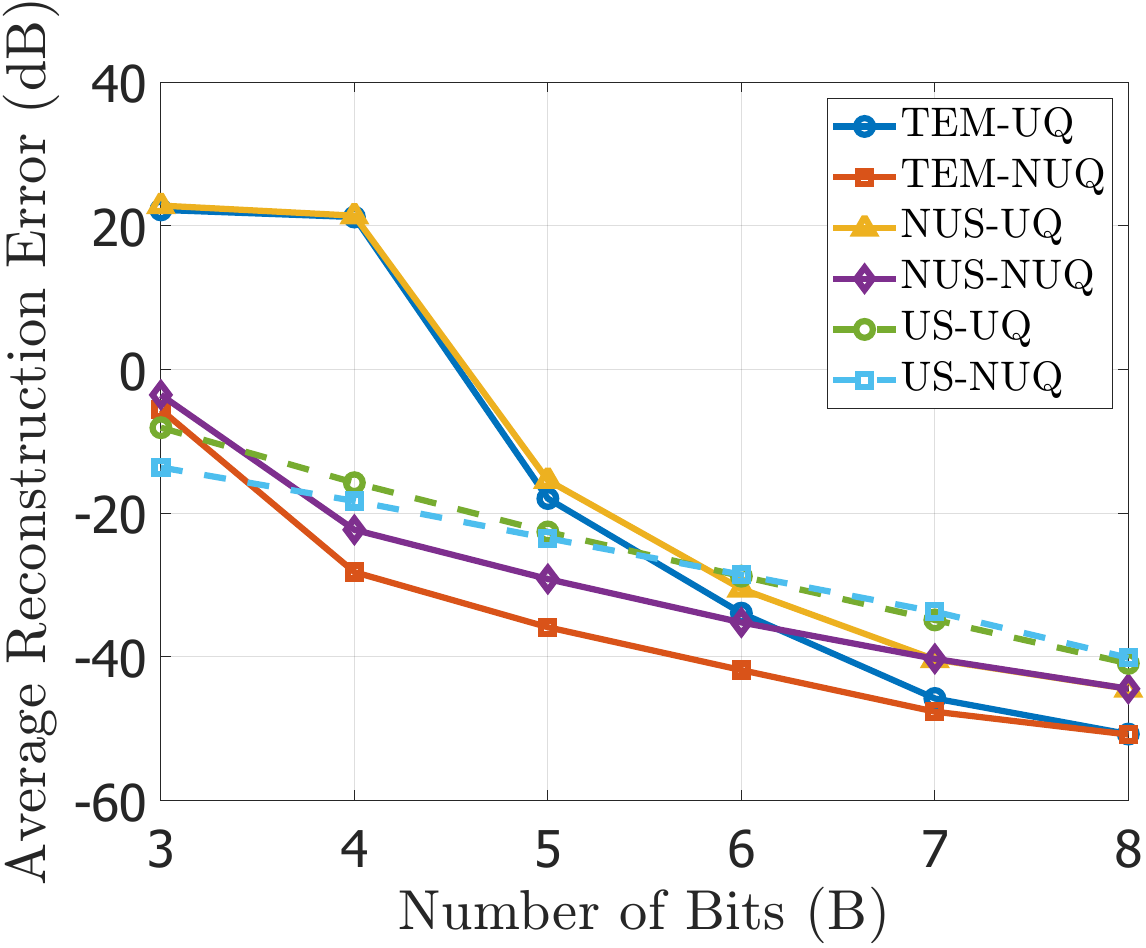}
    \caption{NMSE comparison of TEM and NUS under uniform and non-uniform quantization.}
    \label{fig:updated_main_figure}
\end{figure}

In this section, we present additional simulation results obtained using practical signals. We also compare the performance of uniform sampling (US), using sinc interpolation, with uniform and non-uniform quantization. The US is performed and reconstructed as follows:
\begin{align}
    f(t) = \sum_{n=-\infty}^{\infty} f(nT_{s}) \ \text{sinc}\left(\frac{t - nT_{s}}{T_{s}}\right)
    \label{eq:NyquistReconstruction}
\end{align}
where $\text{sinc}(x) = \frac{\sin(\pi x)}{\pi x}$, and $T_s \leq \frac{\pi}{\Omega_0}$.
In Fig.~\ref{fig:updated_main_figure}, we compare the performance of US with amplitude UQ and NUQ, for the same class of signals, as well as the same number of samples, as presented in Section \ref{sec:simresult}. We observe that TEM-NUQ performs significantly better than both these schemes, and we observe a negligible difference between UQ and NUQ for the US scheme. 

\subsection{Practical Application}
In this section, we consider signals from practical applications. Specifically, we analyze an experimentally acquired ultrasonic guided wave signal reported in \cite{Patil2024}, obtained from a steel pipe instrumented with circumferential piezoelectric thickness shear mode transducers. Torsional guided wave responses recorded at the receivers are compared with a baseline signal from a healthy pipe to identify defects through signal deviations. The received signal is used in this work to evaluate the performance of the proposed TEM-NUQ scheme for the ultrasonic guided wave signals.

\begin{figure}[!t]
    \centering    \includegraphics[width=0.8\linewidth]{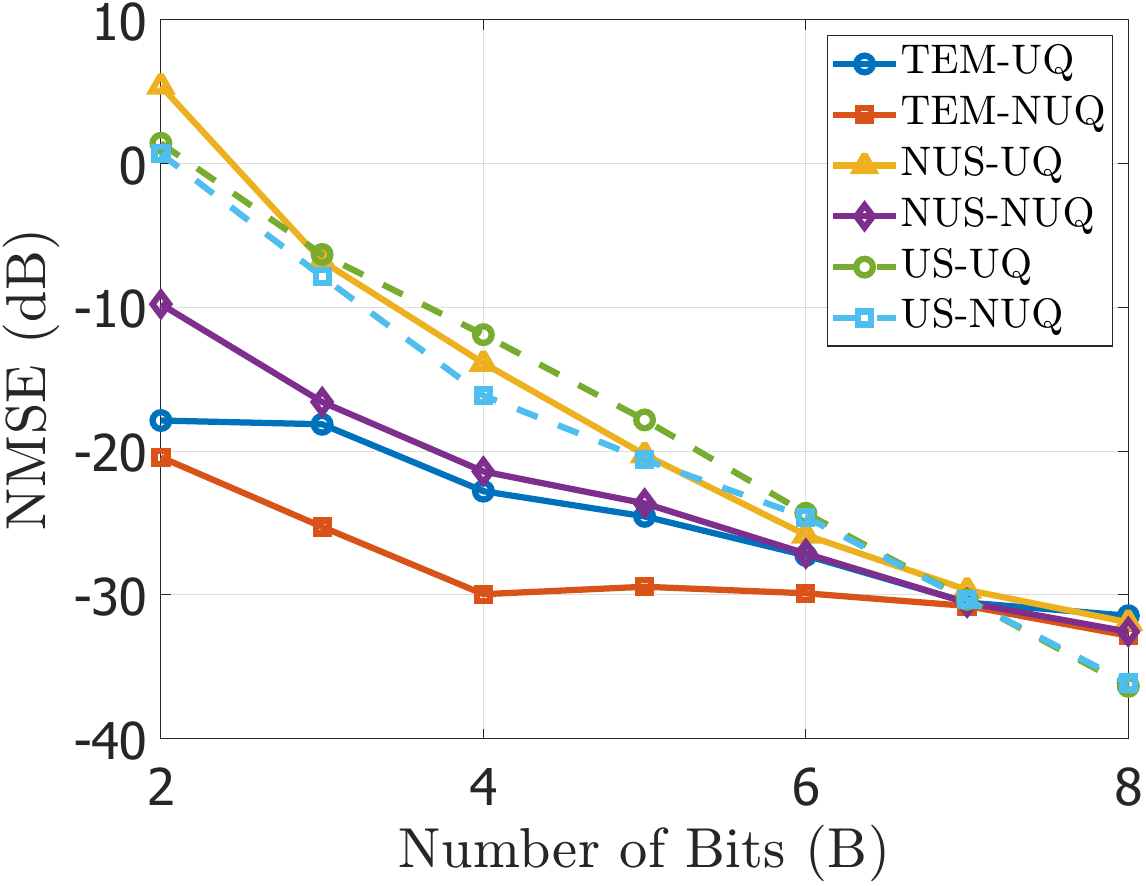}
    \caption{NMSE comparison of TEM, NUS, US under uniform and non-uniform quantization.}
    \label{fig:new_main_figure}
\end{figure}

The following results in Fig.~\ref{fig:new_main_figure} depict the reconstruction error with the number of bits, for this signal, compared with the US scheme as well. We observe a better reconstruction error across all three sampling strategies in the case of NUQ, as designed using the transformation presented in Theorem \ref{th:distribution}. We would like to point out that we are using the same number of samples for all three strategies. The parameters used are $b = 1.67, \Delta = 3.03\times10^{-6}$ for the TEM strategy, and the sampling frequency $F_s = 6 \times 10^{5}$ Hz. 

 \begin{figure}[!t]
    \centering
    \begin{subfigure}{0.48\linewidth}
        \centering
        \includegraphics[width=\linewidth]{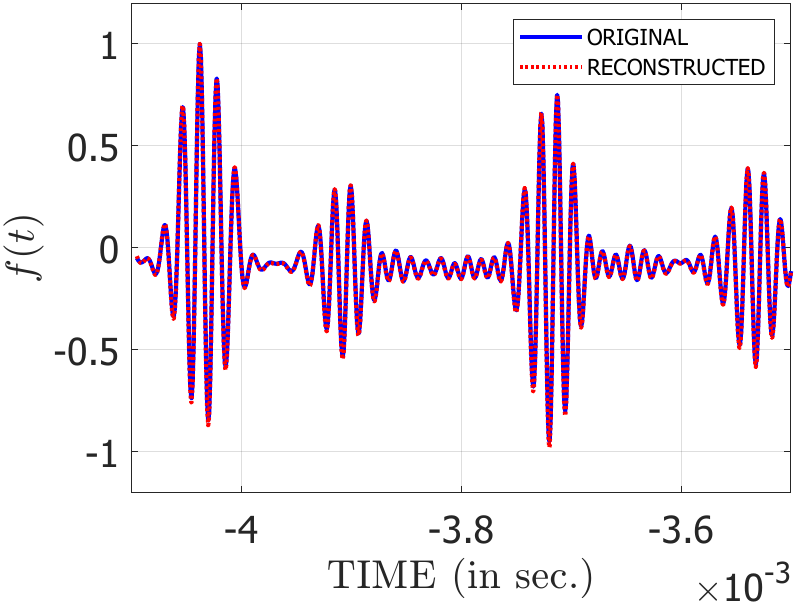}
        \caption{Non-Uniform quantization in TEM with 4 bits ($\text{NMSE} = -29.952 \,\mathrm{dB}$).}
        \label{fig:TEM_error_3}
    \end{subfigure}
    \hfill
    \begin{subfigure}{0.48\linewidth}
        \centering
        \includegraphics[width=\linewidth]{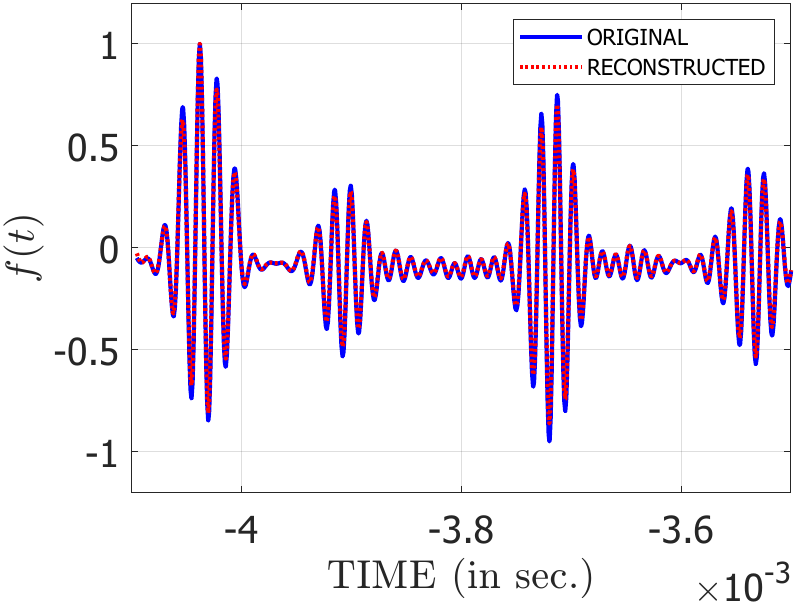}
        \caption{Uniform  quantization in TEM with 4 bits ($\text{NMSE} = -22.789 \,\mathrm{dB}$).}
        \label{fig:TEM_error_4}
    \end{subfigure}

    \caption{Reconstruction error comparison between TEM-NUQ and TEM-UQ for the ultrasonic guided waves.}
    \label{fig:TEM_error_time_UQ}
\end{figure}

As depicted in Fig.~\ref{fig:TEM_error_time_UQ}, at a quantization resolution of $4$ bits, TEM-NUQ achieves the lowest reconstruction error among all considered schemes, providing an improvement of approximately $7$~dB compared to TEM-UQ. 
Here, TEM-NUQ employs non-uniformly quantized time encodings matched to the signal statistics, whereas TEM-UQ uses uniformly quantized time encodings.
These results validate the effectiveness of the proposed NUQ design on practical ultrasonic guided wave signals.

\end{document}